\documentclass[pre,aps,twocolumn,superscriptaddress]{revtex4-1}

\usepackage{graphicx}
\usepackage{amssymb,amsfonts,amsmath}
\usepackage{color}
\usepackage{ulem}
\usepackage[hidelinks]{hyperref}
\usepackage{bbm}
\usepackage{bbold}
\usepackage{algpseudocode}
\usepackage{algorithm}

\usepackage{tikz}
\usetikzlibrary{shapes}
\usetikzlibrary{patterns}
\usetikzlibrary{angles, quotes}

\newcommand{\rv}{{\mathbf r}}

\newcommand{\Tr}{{\rm Tr}\,}
\newcommand{\e}{{\rm e}}

\newcommand{\Jv}{{\bf J}}

\newcommand{\pv}{{\bf p}}

\newcommand{\Fv}{{\bf F}}

\newcommand{\msphantom}[1]{$\ldots$}

\newcommand{\eps}{{\boldsymbol \epsilon}}

\newcommand{\eqr}[1]{Eq.~\eqref{#1}}

\newcommand{\unity}{{\mathbbm 1}}

\newcommand{\mydelete}[1]{{}}
\newcommand{\taub}{{\boldsymbol\tau}}

\newcommand{\rmint}{{\rm int}}

\newcommand{\rmext}{{\rm ext}}

\newcommand{\rmid}{{\rm id}}
\newcommand{\bsig}{\boldsymbol\sigma}

\newcommand{\rmcl}{{\rm cl}}
\newcommand{\Sv}{{\bf S}}

\newcommand{\calO}{{\cal O}}
\newcommand{\rmi}{{\rm i}}

\newcommand{\Hhat}{\hat H}

\renewcommand{\rho}{{n}}

\begin{document}

\title{Quantum statistical mechanical gauge invariance}

\author{Johanna M\"uller}
\affiliation{Theoretische Physik II, Physikalisches Institut, 
  Universit{\"a}t Bayreuth, D-95447 Bayreuth, Germany}

\author{Matthias Schmidt}
\affiliation{Theoretische Physik II, Physikalisches Institut, 
  Universit{\"a}t Bayreuth, D-95447 Bayreuth, Germany}
\email{Matthias.Schmidt@uni-bayreuth.de}

\date{24 September 2025, revised version: 15 July 2026}

\begin{abstract}
We address gauge invariance in the statistical mechanics of quantum
many-body systems. The gauge transformation acts on the position and
momentum degrees of freedom and it is represented by a quantum
shifting superoperator that maps quantum observables onto each
other. The shifting superoperator is anti-self-adjoint and it has
noncommutative Lie algebra structure. These properties induce exact
equilibrium sum rules that connect locally-resolved force and
hyperforce densities for any given observable. We argue that the
framework is amenable to tight integration into quantum hyperdensity
functional theory and that it generalizes naturally to nonequilibrium.
\end{abstract}

\maketitle

The systematic treatment of gauge invariance is key to relating the
symmetries that are inherent in a physical theory to the validity of
exact identities. Typically such equations have the form of
conservation laws that restrict the behaviour of the fundamental
physical degrees of freedom, which are often taken to be fields.  The
gauge transformations can carry intricate mathematical structure, have
profound consequences for our understanding of nature, and they reside
at the core of important modern developments in theoretical
physics~\cite{raifeartaigh2000, jackson2001}.

As a central tool to analyze invariances, Noether's theorem
\cite{noether1918} was used in various different statistical
mechanical settings \cite{baez2013markov, marvian2014quantum,
  sasa2016, sasa2019, bravetti2023, budkov2022, brandyshev2023,
  beyen2025clausius}. A specific `shifting' operation was argued to
constitute a gauge transformation for classical statistical mechanics
in equilibrium \cite{mueller2024gauge, mueller2024whygauge,
  rotenberg2024spotted, miller2025physicsToday} and under general
Hamiltonian dynamics \cite{mueller2024dynamic}. The shifting is a
canonical transformation both in classical \cite{hermann2021noether,
  robitschko2024any} and quantum mechanical form
\cite{hermann2022quantum}. The classical framework leads to force and
generalized `hyperforce' correlation functions that satisfy exact sum
rules \cite{robitschko2024any, sammueller2023whatIsLiquid,
  mueller2024gauge, mueller2024whygauge} and it allows one to
construct and test novel sampling schemes \cite{mueller2024gauge,
  rotenberg2024spotted}.  As a special case the Yvon-Born-Green
equation \cite{yvon1935, born1946, hansen2013}, which expresses the
position-resolved equilibrium force density balance, follows and it is
generalized to a dynamical `hypercurrent' identity
\cite{mueller2024dynamic}.

Here we present the generalization of the classical gauge invariance
to quantum many-body systems; consequences will be presented in
Ref.~\cite{mueller2025longQuantum}. We demonstrate that all salient
features of the classical gauge theory remain intact, including the
validity of exact static hyperforce and dynamical hypercurrent sum
rules, see \eqr{EQhyperForceDensityBalance}, provided that the
definitions of the occuring quantum correlation functions are
appropriately identified, which is nontrivial. The perseverance of the
theoretical structure is remarkable, given the fundamental changes in
the underlying microscopic description of the many-body physics.

Quantum statistical mechanics applies to a broad range of diverse
systems of current interest. Topical examples include ultracold atomic
systems \cite{bloch2008}, where questions of many-body localization
are pertinent from both experimental \cite{choi2016, bordia2017} and
theoretical \cite{yan2017prl,yan2017pra} perspectives.
Systems that were initally thermalized and then undergo a sudden
change via explicit time dependence of a tuning parameter in the
Hamiltonian constitute an important class of setups.  Whether then
thermalization occurs at long times is a relevant question, see
e.g.~Refs.~\cite{lydzba2023, patil2026} for recent work and
Ref.~\cite{patil2026review} for an introduction.  Much attention is
being paid to the study of spatially non-homogeneous quantum many-body
systems \cite{palamara2024,palamara2025,ullrich2025}.

When working with discrete particles instead of fields, then a quantum
many-body description involves the position and momentum degrees of
freedom of each particle. The quantum nature of the problem is
reflected by the algebraic commutator structure of the canonical
quantum operators.  Formulating a reduced picture can be based
efficiently on the density operator $\hat\rho(\rv)=\sum_i
\delta(\rv-\rv_i)$, see e.g.~Ref.~\cite{schmidt2022rmp}, where the sum
runs over all particles $i=1,\ldots,N$, the variable $N$ is the total
number of particles, $\delta(\cdot)$ denotes the Dirac distribution in
$d$ dimensions, $\rv_i$ is the position of particle $i$, and $\rv$ is
a generic position.

When considering a general quantum observable $\hat A $, the quantum
dynamics are given by the associated Heisenberg equation of motion,
$\partial \hat A(t)/\partial t = (-\rmi/\hbar)[\hat A(t),\Hhat(t)]$, where
$[\,\cdot\,,\,\cdot\,]$ denotes the commutator, $\Hhat$ is the
Hamiltonian, $\hbar$~indicates the reduced Planck constant, $\rm i$ is
the imaginary unit, $t$ denotes time, and we let $\hat A$ be
stationary, such that it carries no mere parametric dependence on
time.  Choosing the density operator as the observable of interest,
$\hat A=\hat \rho(\rv)$, the generic Heisenberg equation of motion
reduces to the continuity equation, $\partial \hat\rho(\rv,t)/\partial
t = -\nabla\cdot \hat\Jv(\rv,t)$, where $\nabla$ indicates the
derivative with respect to $\rv$. Here $\hat\Jv(\rv,t)$ is the
one-body current operator (defined below). Its dynamics follow again
from Heisenberg's equation of motion, which yields $\partial
m\hat\Jv(\rv,t)/\partial t=\hat\Fv(\rv,t)$, where the result is the
one-body force density operator (also described below) and $m$
indicates the particle mass.  Hence one has arrived at a spatially
resolved analog of Newton's second law and upon building the quantum
average one recovers Ehrenfest's theorem, again in a spatially
resolved version.  The crucial step in the derivation is to obtain a
microscopic expression for the force density operator
\cite{schmidt2022rmp} by calculating explicitly the commutator
$(-\rmi/\hbar)[m\hat\Jv(\rv,t),\Hhat(t)] = \hat \Fv(\rv,t)$. This
formulation forms a perfectly valid starting point for developing the
quantum statistical mechanics of $N$-body systems.
A description of the relationship with classical statistical mechanics
is given in Ref.~\cite{mueller2025longQuantum}.

Here we formulate {\it quantum} shifting gauge invariance.  Inspired
by the recent classical formulation of gauge shifting based on Poisson
brackets \cite{mueller2024dynamic} we here introduce the following
quantum `shifting superoperator':
\begin{align}
  \bsig(\rv) &= -\frac{i}{\hbar}[\,\cdot\,,m\hat \Jv(\rv)],
  \label{EQqsigDefinition}
\end{align}
where the (scaled) quantum one-body current (density) operator in
Schr\"odinger form is:
\begin{align}
  m\hat\Jv(\rv) &= \frac{1}{2}\sum_i[\hat\pv_i\delta(\rv-\rv_i) +
    \delta(\rv-\rv_i)\hat\pv_i],
  \label{EQcurrentDefinition}
\end{align}
with $\hat\pv_i = -\rmi \hbar \nabla_i$ denoting the momentum operator
of quantum particle~$i$, where $\nabla_i = \partial / \partial \rv_i$.
Equation \eqref{EQqsigDefinition} defines a quantum superoperator, as
this applies to Hilbert space operators (first argument of the
commutator) and it also returns a Hilbert space operator (the result
of the scaled commutator).  We demonstrate in the following that the
shifting superoperator \eqref{EQqsigDefinition} is the appropriate
quantum mechanical entity to encapsulate statistical mechanical gauge
invariance by laying out several of its key properties. 

As an aside, the relationship with classical phase space
shifting \cite{mueller2024dynamic}, as based on Dirac's correspondence
principle \cite{dirac1958book}, is 
noteworthy. In the classical description of the many-body statistical
mechanics, the shifting acts on phase space \cite{mueller2024gauge,
  mueller2024whygauge}. The differential operators that generate the
transform can be expressed~\cite{mueller2024dynamic} based on Poisson
brackets as $\bsig_\rmcl(\rv)=\{\,\cdot\,, m\hat\Jv_\rmcl(\rv)\}$,
with $\{\,\cdot\,,\,\cdot\,\}$ denoting the (classical) Poisson
brackets and $\hat\Jv_\rmcl(\rv)$ the classical phase space current
observable.  We note the analogy of $\bsig_\rmcl(\rv)$ to the quantum
shifting superoperator~\eqref{EQqsigDefinition} and refer the Reader
to Ref.~\cite{mueller2025longQuantum} for a detailed description of
the (nontrivial) classical-quantum shifting correspondence.

As a seemingly trivial initial case, applying the quantum shifting
superoperator \eqref{EQqsigDefinition} to the identity operator~1
yields
\begin{align}
  \bsig(\rv)1 &= 0,
  \label{EQqsigIdentiy}
\end{align}
which follows from the definition \eqref{EQqsigDefinition} and the
identity commuting with the current operator, $[1, m\hat\Jv(\rv)]=0$.
The application of the superoperator to the negative Hamiltonian
yields the force density operator,
\begin{align}
  \hat\Fv(\rv) &=  -\bsig(\rv) \Hhat,
  \label{EQqsigHamiltonian} 
\end{align}
which constitutes the Schr\"odinger form of $\hat\Fv(\rv,t)$ and is
specified in more detail later.  Equation \eqref{EQqsigHamiltonian}
follows from the definition \eqref{EQqsigDefinition} and the
Heisenberg equation of motion for the current operator.

When considering a general quantum observable $\hat A$ and applying
$\bsig(\rv)$ we refer to the result as the {\it hyperforce} density
operator
\begin{align}
  \hat\Sv_A(\rv) = \bsig(\rv) \hat A,
  \label{EQqsigHyperforce}
\end{align}
which is the quantum analog of the corresponding classical phase space
function \cite{mueller2024gauge, robitschko2024any} and is expressed
as $\hat\Sv_A(\rv)=(-\rmi/\hbar)[\hat A, m\hat\Jv(\rv)]$ according to
the commutator form~\eqref{EQqsigDefinition}.  That $\hat \Sv_A(\rv)$
is indeed a quantum observable,
$\hat\Sv_A^\dagger(\rv)=\hat\Sv_A(\rv)$, follows from $[\bsig(\rv)\hat
  A]^\dagger = \bsig(\rv)\hat A$ for $\hat A = \hat A^\dagger$. This
property can in turn be derived from
\begin{align}
  [\bsig(\rv)\hat A]^\dagger = \bsig(\rv)\hat A^\dagger,
  \label{EQqsigAadijoint}
\end{align}
where $\hat A$ can be general (not necessarily self-adjoint).
Equation~\eqref{EQqsigAadijoint} is a consequence of the commutator
structure~\eqref{EQqsigDefinition} and the self-adjointness of the
current operator, $m\hat \Jv(\rv) = m\hat \Jv^\dagger(\rv)$.

For completeness, adjoining an operator $\hat A$ is indicated by the
dagger, $\hat A^\dagger$, and defined throughout in the standard way
\cite{dirac1958book, sakurai1973book} as $ \langle n'|\hat A n\rangle=
\langle \hat A^\dagger n'|n\rangle$ for all $n$ and $n'$ of a general
$N$-body Hilbert space basis $|n\rangle$. By extension
\cite{fick1990book}, the adjoint ${\cal O}^\dagger$ of a general
superoperator~$\cal O$ is defined as $\Tr \hat A^\dagger {\cal O} \hat
B= \Tr ({\cal O}^\dagger \hat A)^\dagger \hat B$, where $\hat A$ and
$\hat B$ are general Hilbert space operators and $\Tr\,\cdot\, =
\sum_n \langle n | \,\cdot \,|n\rangle$ indicates the trace over
Hilbert space. This will be important for the statistical mechanics
described later.

In this extended sense, the shifting superoperator
\eqref{EQqsigDefinition} is anti-self-adjoint,
\begin{align}
  \bsig^\dagger(\rv) &= -\bsig(\rv),
  \label{EQqsigAntiSelfAdjoint}
\end{align}
which follows from the quantum trace being invariant under cyclic
permutations~\cite{fick1990book}.  More specifically, $\Tr \hat A[\hat
  B, \hat C] = \Tr [\hat C, \hat A] \hat B $, where one chooses $\hat
C = (-\rmi/\hbar)m\hat\Jv(\rv)$. Together with \eqr{EQqsigDefinition}
this leads to
\begin{align}
  \Tr \hat A [\bsig(\rv) \hat B]
  &= -\Tr [\bsig(\rv)\hat A] \hat B,
  \label{EQqsigTraceCyclic}
\end{align}
where replacing $\hat A$ with $\hat A^\dagger$ and combining with
\eqr{EQqsigAadijoint} gives \eqr{EQqsigAntiSelfAdjoint}.

While the relationships
\eqref{EQqsigIdentiy}--\eqref{EQqsigTraceCyclic} point towards the
prowess of individual uses of the shifting superoperator, one needs to
consider multiple instances of $\bsig(\rv)$ to reveal the full
mathematical structure.  Before we demonstrate below the existence of
a Lie superoperator algebra, we first address the commutator of two
shifting superoperators:
\begin{align}
  [\bsig(\rv), \bsig(\rv')] &=
  [\nabla\delta(\rv-\rv')]\bsig(\rv)
  +\bsig(\rv')[\nabla\delta(\rv-\rv')],
  \label{EQqsigCommutator}
\end{align}
where $[\nabla\delta(\rv-\rv')]$ is the derivative of the Dirac
distribution and the brackets limit the scope of $\nabla$. Analogously
one can express $\nabla\delta(\rv-\rv')=(\rmi/\hbar)[\hat\pv,
  \delta(\rv-\rv')]$ upon introducing a generic momentum operator
$\hat\pv = -\rmi \hbar \nabla$ that satisfies the canonical commutator
relationship $[\rv,\hat\pv]=\rmi \hbar \unity$ with the generic
position $\rv$. The commutator of two superoperators $\calO_1$ and
$\calO_2$ is defined as $[\calO_1, \calO_2] \hat A = \calO_1 (\calO_2
\hat A) - \calO_2 (\calO_1\hat A)$ for any general operator $\hat A$.

To connect to the quantum shifting transform \cite{hermann2022quantum}
and to be able to identify the Lie superoperator algebra we consider
the integrated shifting superoperators $\Sigma[\eps] = \int d\rv
\eps(\rv)\cdot\bsig(\rv)$, where the square brackets indicate
functional dependence on the (smooth) shifting field $\eps(\rv)$. An
explicit form of $\Sigma[\eps]$ is obtained by expressing $\bsig(\rv)$
via the commutator~\eqref{EQqsigDefinition}, using the scaled current
\eqref{EQcurrentDefinition}, and carrying out the position integral,
which yields:
\begin{align}
  \Sigma[\eps] &=
  -\frac{\rmi}{2\hbar}
  \sum_i[\,\cdot\,, \eps(\rv_i)\cdot\hat\pv_i
    +\hat\pv_i\cdot\eps(\rv_i)].
  \label{EQSigmaOfEpsDefinition}
\end{align}

As an illustration of \eqr{EQSigmaOfEpsDefinition} we consider the
effect of $1+\Sigma[\eps]$ acting on the fundamental position and
momentum degrees of freedom. The result is
$(1+\Sigma[\eps])\rv_j = \rv_j + \eps(\rv_j)$ and
$(1+\Sigma[\eps])\hat\pv_j = \hat\pv_j -
[(\nabla_j\eps(\rv_j))\cdot\hat\pv_j + \hat\pv_j \cdot (\nabla_j
  \eps(\rv_j))^{\sf T}]/2$, where the superscript $\sf T$ indicates
transposition of a $d\times d$ matrix. Thus we recover the
(linearized) quantum canonical shifting transformation of
Ref.~\cite{hermann2022quantum}, where quantum canonical
transformations \cite{anderson1994} generalize their classical
counterparts \cite{goldstein2002}. 

Working with the superoperator \eqref{EQqsigDefinition} allows one to
make significant progress over the results obtained via the explicit
quantum canonical transformation \cite{hermann2022quantum}, as we
demonstrate in the following. Given two smooth shifting fields
$\eps_1(\rv)$ and $\eps_2(\rv)$ the corresponding integrated
superoperators form a noncommutative Lie algebra,
\begin{align}
  [\Sigma[\eps_1], \Sigma[\eps_2]] = \Sigma[\eps_\Delta].
  \label{EQSigmaLieAlgebra}
\end{align}
The explicit form of the difference shifting field is
\begin{align}
  \eps_\Delta(\rv) &= \eps_1(\rv)\cdot\nabla\eps_2(\rv) 
  - \eps_2(\rv)\cdot\nabla\eps_1(\rv),
  \label{EQepsDelta}
\end{align}
which is identical to the standard Lie bracket of the two vector
fields $\eps_1(\rv)$ and $\eps_2(\rv)$, as holds also
classically~\cite{mueller2024gauge, mueller2024whygauge}.  Using the
generic momentum operator $\hat\pv$ allows one to express
\eqr{EQepsDelta} alternatively as $\eps_\Delta(\rv) =
(\rmi/\hbar)\big(\eps_1(\rv)\cdot[\hat\pv, \eps_2(\rv)] -
\eps_2(\rv)\cdot[\hat\pv, \eps_1(\rv)] \big)$.  The derivation of
\eqr{EQSigmaLieAlgebra} is based on resolving the nested commutators
on the left hand side.  The commutator relationship
\eqref{EQqsigCommutator} between the localized superoperators
$\bsig(\rv)$ and $\bsig(\rv')$ then follows via building the mixed
second functional derivative of \eqr{EQSigmaLieAlgebra},
$\delta^2/[\delta\eps_1(\rv)\delta\eps_2(\rv')]$, observing that
$\delta \Sigma[\eps]/\delta\eps(\rv)= \bsig(\rv)$, and simplifying.

Having laid out the geometrical structure of the quantum mechanical
shifting, we turn to its statistical mechanical consequences. We
consider the following generic many-body Hamiltonian
\begin{align}
  \Hhat &= \sum_i \frac{\hat \pv_i^2}{2m}
  + u(\rv^N) + \sum_i V_\rmext(\rv_i),
  \label{EQHamiltonian}
\end{align}
where $u(\rv^N)$ is the interparticle interaction potential and
$V_\rmext(\rv)$ is an external one-body potential. We consider
stationary Hamiltonians $\Hhat_0$, where the subscript~0 indicates the
absence of explicit time dependence. The corresponding canonical
quantum partition sum is $ Z = \Tr \e^{-\beta \Hhat_0}$, where
$\beta=1/(k_BT)$ with Boltzmann constant $k_B$ and absolute
temperature~$T$. The canonical free energy is $F=-k_BT\ln Z$ and
thermal equilibrium averages of general quantum observables are given
by $\langle \,\cdot\, \rangle = \Tr \,\cdot\, \e^{-\beta \Hhat_0}/Z$.
This formulation encompasses both bosonic and fermionic systems. The
respective exchange symmetry characterizes the Hilbert space basis
$|n\rangle$, such that $|n\rangle$ acquires a factor $+1$ (bosons) or
$-1$ (fermions) upon particle exchange; details are provided in
Ref.~\cite{mueller2025longQuantum}.

One defining feature of any gauge transformation is that its
application leaves measurable quantities invariant, which in the
present case are quantum statistical mechanical averages.  Applying
the integrated shifting superoperator~\eqref{EQSigmaOfEpsDefinition}
to a given observable $\hat A$ will in general have a nonvanishing
effect on $\hat A$, such that $(1+\Sigma[\eps])\hat A = \hat A +
\Sigma[\eps] \hat A \neq \hat A$, since $\Sigma[\eps] \hat A\neq
0$. However, on average $\langle \Sigma[\eps]\hat A \rangle= 0$,
irrespective of the specific form of the gauge function $\eps(\rv)$
and of the observable $\hat A$; we recall the thermal mean as $\langle
\Sigma[\eps] \hat A \rangle= \Tr \Sigma[\eps]\hat A \e^{-\beta
  \Hhat_0}/Z$ with $\Sigma[\eps]$ acting on the product $\hat
A\e^{-\beta \Hhat_0}$. That the average vanishes can be seen by
expressing $\Sigma[\eps]$ in the form given above
\eqr{EQSigmaOfEpsDefinition} to re-write $\langle \Sigma[\eps] \hat A
\rangle = \int d\rv \eps(\rv)\cdot\langle \bsig(\rv) \hat A \rangle$,
where $\langle \bsig(\rv)\hat A \rangle=0$ due
to $\langle \bsig(\rv)\hat A \rangle = \langle
[\bsig^\dagger(\rv)1]^\dagger\hat A \rangle = - \langle
[\bsig(\rv)1]^\dagger\hat A \rangle =0$, as follows from the
anti-self-adjointness~\eqref{EQqsigAntiSelfAdjoint} and
\eqr{EQqsigIdentiy}.

The localized shifting superoperator \eqref{EQqsigDefinition} has no
dependence on the gauge function and hence it forms a very efficient
starting point for the derivation of sum rules. As a prerequisite, we
consider applying the shifting superoperator to the Boltzmann factor:
\begin{align}
  \bsig(\rv) \e^{-\beta \Hhat_0} &= \int_0^\beta d\beta'
  \e^{-\beta' \Hhat_0} \hat\Fv_0(\rv) \e^{\beta'\Hhat_0} \e^{-\beta \Hhat_0},
  \label{EQqsigBoltzmannfactor}
\end{align}
where the force density operator \eqref{EQqsigHamiltonian} is
\begin{align}
  \hat\Fv_0(\rv) &= -\bsig(\rv) \Hhat_0.
  \label{EQqsigInitialStateHamiltonian}
\end{align}
Equation \eqref{EQqsigBoltzmannfactor} follows from the commutator
form~\eqref{EQqsigDefinition} and the general property of
exponentiated operators $[\hat B, \e^{-\beta \hat C}] = -\int_0^\beta
d\beta'\e^{-\beta'\hat C}[\hat B, \hat C]\e^{\beta' \hat
  C}\e^{-\beta\hat C}$ by setting $\hat B=(\rmi/\hbar) m \hat
\Jv(\rv)$ and $\hat C=\Hhat_0$.  The additive structure of the
Hamiltonian \eqref{EQHamiltonian} induces the force density
splitting~\cite{schmidt2022rmp} into $\hat\Fv_0(\rv) = \nabla\cdot
\hat\taub_0(\rv) + \hat\Fv_{\rmint,0}(\rv) - \hat\rho(\rv) \nabla
V_{\rmext,0}(\rv)$, where $\hat\taub_0(\rv)$ is the kinetic stress
tensor \cite{schmidt2022rmp, hermann2022quantum},
$\hat\Fv_{\rmint,0}(\rv) = -\sum_i\delta(\rv-\rv_i)[\nabla_i
  u_0(\rv^N)]$ is the interparticle force density, and
$V_{\rmext,0}(\rv)$ is the external potential, all for the equilibrium
system.

To start the sum rule construction, we first build the equilibrium
average of the trivial \eqr{EQqsigIdentiy}. This yields $0 = \langle 0
\rangle = \langle [\bsig(\rv) 1]^\dagger \rangle = \langle
\bsig^\dagger(\rv) \rangle = -\langle \bsig(\rv) \rangle = -\Tr
\bsig(\rv) \e^{-\beta \Hhat_0} /Z = -\langle \beta\hat\Fv_0(\rv)
\rangle$, which follows from \eqr{EQqsigAntiSelfAdjoint}, writing out
the canonical average, using the Boltzmann operator identity
\eqref{EQqsigBoltzmannfactor}, using the invariance of the trace under
cyclic permutations, and identifying the thermal average. Defining the
mean force density as $\Fv_0(\rv)=\langle \hat \Fv_0(\rv)\rangle$
leads to
\begin{align}
  \Fv_0(\rv) &= 0,
  \label{EQforceDensityBalance}
\end{align}
which is the equilibrium force density balance \cite{schmidt2022rmp,
  hermann2022quantum}.

\begin{figure}[tbh!]
  \includegraphics[page=1,width=.92\columnwidth]{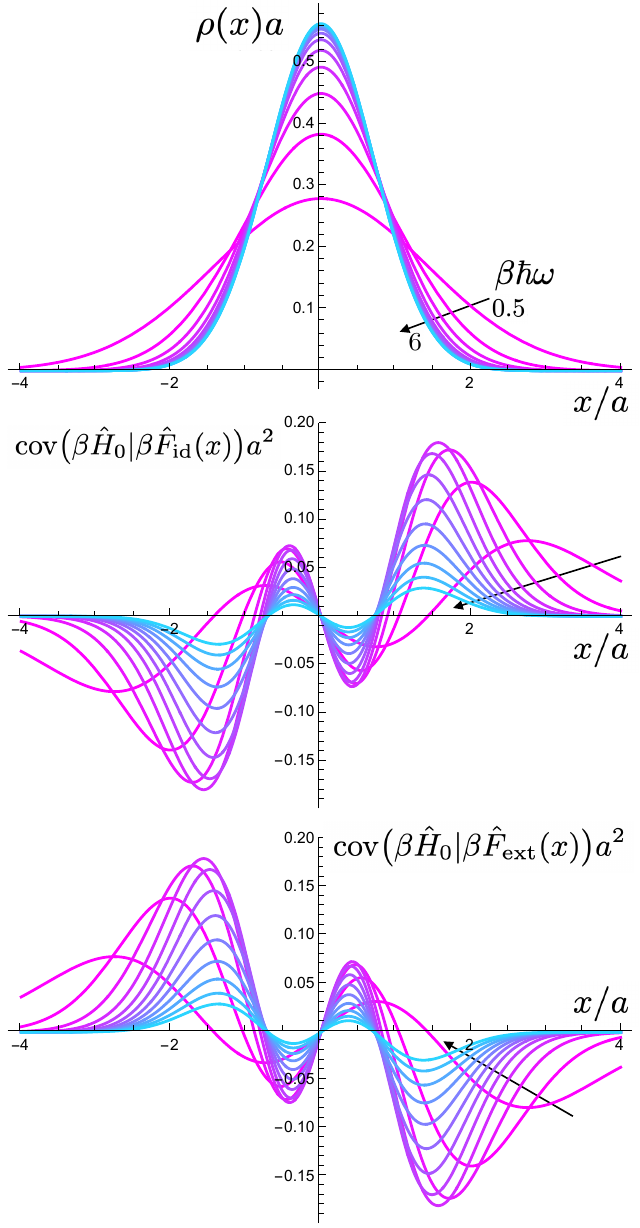}
  \caption{Demonstration of the general sum rule
    \eqref{EQhyperForceDensityBalance} applied to the observable $\hat
    A = \beta \Hhat_0$ of a harmonic oscillator with frequency
    $\omega$. The results are for different values of scaled inverse
    temperature $\beta\hbar\omega=0.5, 1, \ldots, 6$ (from magenta to
    blue) and shown as a function of the scaled coordinate $x/a$ with
    lengthscale $a=\sqrt{{\textcolor{black}{\hbar/(m
          \omega)}}}$. Shown are thermal density profile $\rho(x)a$
    (top panel), kinetic Mori term ${\rm cov}\big( \beta \Hhat_0
    |\beta \hat F_\rmid(x) \big) a^{2}$ (middle panel), and external
    force Mori term ${\rm cov}\big( \beta \Hhat_0 | {\beta}\hat
    F_\rmext(x) \big) a^{2}$ (bottom panel).  Both Mori covariances
    display pronouced spatial structuring. For each value of~$\beta$
    the sum of the ideal and external covariances vanishes.  In
    general the interparticle covariance term will contribute.  Here
    the general force density operator \cite{schmidt2022rmp} has the
    ideal contribution $\hat F_\rmid(x) = \partial
    \hat\tau(x)/\partial x$, with one-dimensional kinetic stress
    operator $\hat\tau(x)=\hbar^2/(4m) \partial^2 \hat n(x)/\partial
    x^2-\sum_i\hat p_i\delta(x-x_i)\hat p_i/m$, density operator $\hat
    n(x)=\sum_i\delta(x-x_i)$, external contribution $\hat F_\rmext(x)
    = -m\omega^2x \hat n(x)$, and fundamental degrees of freedom
    $x_i$, $\hat p_i=-\rmi \hbar \partial/\partial x_i$ with $N=1$ in
    the present case.}
\label{FIG1}
\end{figure}

To incorporate general observables $\hat A$ into the framework, we
apply the averaging strategy to the adjoint of \eqr{EQqsigHyperforce}.
On the left hand side this yields $\langle
\hat\Sv_A^\dagger(\rv)\rangle = \langle \hat\Sv_A(\rv) \rangle =
\Sv_A(\rv)$, where we have used the self-adjointness of the hyperforce
density operator and then have defined the mean hyperforce density
$\Sv_A(\rv)$.  On the right hand side one obtains $\langle[ \bsig(\rv)
  \hat A]^\dagger \rangle = \langle \hat A^\dagger \bsig^\dagger(\rv)
\rangle = - \langle \hat A^\dagger \bsig(\rv) \rangle = - \Tr \hat
A^\dagger \bsig(\rv) \e^{-\beta \Hhat_0}/Z = -\big(\hat A| \beta\hat
\Fv_0(\rv)\big)$, where we have first built the adjoint of the
shifting superoperator, used its
anti-self-adjointness~\eqref{EQqsigAntiSelfAdjoint}, and then written
out the thermal average. In the last step we have first applied
\eqr{EQqsigBoltzmannfactor} and in the notation used the
Mori(-Kubo-Bogoliubov) product $(\,\cdot\,|\,\cdot\,)$ as a general
means to describe response \cite{fick1990book,kubo1957, petz1993}.
The Mori product constitutes a scalar product of two general operators
$\hat A$ and $\hat B$ and it is defined as $(\hat A | \hat B) =
\beta^{-1} \int_0^\beta d\beta' \Tr \hat A^\dagger \e^{-\beta'\Hhat_0}
\hat B \e^{\beta'\Hhat_0} \e^{-\beta \Hhat_0}/Z$, where here $\hat B =
\beta\hat \Fv_0(\rv)$.  An alternative and equivalent form is $(\hat
A|\hat B) = \beta^{-1}\int_0^\beta d\beta' \langle \hat A^\dagger \hat
B({\rm i}\hbar \beta')\rangle$, with $\hat B({\rm i}\hbar\beta')$
denoting the Heisenberg operator evaluated at imaginary time $t={\rm
  i}\hbar \beta'$, see e.g.\ Ref.~\cite{sauermann1996}. When applied
to the present case we obtain:
\begin{align}
  \big(\hat A | \beta\hat\Fv_0(\rv)\big)
  &= \int_0^\beta d\beta' \langle
  \hat A^\dagger \e^{-\beta'\Hhat_0} \hat\Fv_0(\rv) \e^{\beta'\Hhat_0}
  \rangle,
  \label{EQMoriExplicit}
\end{align}
where we have identified the thermal average $\langle\,\cdot\,\rangle$
on the right hand side.

We recall that \eqr{EQMoriExplicit} is the thermal average of the
adjoint right hand side of \eqr{EQqsigHyperforce}, which equals
$\Sv_A(\rv)$, see above. Re-arranging the equality one obtains the
following equilibrium quantum hyperforce balance:
\begin{align}
  \Sv_A(\rv) + \big( \hat A | \beta\hat\Fv_0(\rv) \big) &= 0,
  \label{EQhyperForceDensityBalance}
\end{align}
which is exact. Since $\hat A = \hat A^\dagger$ both terms in
\eqr{EQhyperForceDensityBalance} are real-valued. Hence one can
express the Mori product alternatively as $(\beta\hat \Fv_0(\rv)|\hat
A)$ or as the Mori covariance ${\rm cov}\big(\hat A | \beta
\hat\Fv_0(\rv) \big) = \big(\hat A | \beta \hat\Fv_0(\rv)) - \langle
\hat A^\dagger \rangle \langle \beta \hat \Fv_0(\rv)\rangle$, where
the latter holds true due to the vanishing mean force
density~\eqref{EQforceDensityBalance}.  The sum rule
\eqref{EQhyperForceDensityBalance} can alternatively be obtained by
considering the forces in an extended ensemble with modified
Hamiltonian, see Ref.~\cite{mueller2025longQuantum} for the
derivation.

As a consistency check, choosing $\hat A=1$ in
\eqr{EQhyperForceDensityBalance} gives from \eqr{EQqsigHyperforce} the
result $\hat\Sv_{\hat A=1}(\rv)=0$, which from
\eqr{EQhyperForceDensityBalance} equals $-\big(1|\beta
\hat\Fv_0(\rv)\big) = -\beta\Fv_0(\rv)$, such that the hyperforce
density balance \eqref{EQhyperForceDensityBalance} reduces to the
equilibrium force density balance~\eqref{EQforceDensityBalance}.  As a
specific example we choose the sum of positions, $\hat A=\sum_i
\rv_i$, which yields from \eqr{EQqsigHyperforce} the hyperforce
density operator as $\hat\Sv_A(\rv)=(-{\rm i}/\hbar)[\sum_i
  \rv_i,m\hat \Jv(\rv)] = \hat\rho(\rv)\unity$, as follows from
explicit calculation of the commutator. Then the sum rule
\eqref{EQhyperForceDensityBalance} yields the density profile:
$\rho(\rv)\unity = -(\sum_i \rv_i|\beta \hat\Fv_0(\rv))$.  For the
choice $\hat A=\beta \Hhat_0$, due to \eqr{EQforceDensityBalance} the
sum rule \eqref{EQhyperForceDensityBalance} attains the form ${\rm
  cov}\big(\beta \Hhat_0|\beta \hat\Fv_0(\rv)\big)=0$. The harmonic
oscillator, as a toy model, is used to exemplify the validity in
Fig.~\ref{FIG1}.

The shifting superoperator can be used in flexible ways and we present
in Ref.~\cite{mueller2025longQuantum} the derivation of general
two-body and product sum rules. While we have worked with fixed number
of particles, all our considerations and resulting sum rules remain
valid in the grand ensemble. One merely needs to replace the canonical
trace with the grand canonical analog, $\Tr'\,\cdot\,=
\sum_{N=0}^\infty\sum_n \langle n|\,\cdot\,|n\rangle e^{\beta\mu
  N}/N!$, where $\mu$ denotes the chemical potential. The gauge theory
applies to both fermions and bosons, as the exchange symmetry is
encoded solely in the nature of the Hilbert space basis~$|n\rangle$.

In conclusion, we have addressed the consequences of invariance
against shifting in quantum many-body systems. That the formal
structure of the resulting quantum gauge theory mirrors closely that
of the classical version \cite{mueller2024gauge, mueller2024whygauge,
  rotenberg2024spotted, miller2025physicsToday, mueller2024dynamic} is
remarkable, given the stark differences between the mathematical
objects that are involved.  The quantum sum rules become formally
analogous to their classical counterparts \cite{mueller2024gauge,
  mueller2024whygauge} upon identifying the quantum operators
$\hat\Sv_A(\rv)$ and $\hat A$ with the respective classical phase
space functions and reducing the Mori product to the thermal average
of the classical phase space product. We provide further details and a
description of how the analogy relates to Dirac's correspondence
principle \cite{dirac1958book} in Ref.~\cite{mueller2025longQuantum}.

Due to its applicability to general observables, the hyperforce sum
rule \eqref{EQhyperForceDensityBalance} provides much potential for
the integration within further theoretical approaches.  As a
demonstration of such uses, we present in
Ref.~\cite{mueller2025longQuantum} the quantum version of hyperdensity
functional theory \cite{sammueller2024hyperDFT,
  sammueller2024whyhyperDFT}, which provides a framework to represent
the thermal equilibrium behaviour of general observables as density
functionals \cite{mermin1965}.  Dynamical situations in which the
initial thermal system with Hamiltonian $\Hhat_0$ is driven out of
equilibrium by a general time-dependent Hamiltonian $\Hhat$ are also
addressed in Ref.~\cite{mueller2025longQuantum}. In this
nonequilibrium setup the dynamical gauge invariance yields an exact
`hypercurrent' sum rule \cite{mueller2025longQuantum}, which provides
a nonequilibrium generalization of the thermal hyperforce balance
\eqref{EQhyperForceDensityBalance} and is the quantum analog of the
corresponding classical result~\cite{mueller2024dynamic}.

In future work, it would be interesting to further explore connections
with modern developments in density functional theory
\cite{tokatly2005one, tokatly2005two, tokatly2007, ullrich2006,
  tarantino2021, tchenkoue2019, palamara2024, palamara2025, daas2025,
  ullrich2025} and with standard approaches, such as linear response
and the Green-Kubo theory \cite{green1954, kubo1957} and quantum work
relations \cite{deroeck2004}. Much inspiration could come from the
recent gauge investigations of quantum
thermodynamics~\cite{celeri2024, ferrari2025}, from the treatment of
orientational degrees of freedom in classical anisotropic fluids
\cite{nguyen2026}, from the relationship with standard symmetries as
recently investigated classically~\cite{phamvan2026symmetry}, as well
as from distributional formulations within rigorous functional
analysis \cite{maruyama2026}. We also refer the Reader to
  Pham-Van's deep very recent investigations into quantum statistical
  gauge geometry \cite{phamvan2026quantumGeometry,
    phamvan2026exchangePrivateLindblad}.

\smallskip

We thank Florian Samm\"uller, Robert Evans, and Hai Pham-Van for
useful discussions.  This work is supported by the DFG (Deutsche
Forschungsgemeinschaft) under Project No.~551294732.

\smallskip

Data Availability Statement.  The data that support the findings of
this article are openly available~\cite{mueller2025quantumData}.


\begin{thebibliography}{10}

\bibitem{raifeartaigh2000}
L. O'Raifeartaigh and N. Straumann, Gauge theory: Historical origins and some
  modern developments, \href{https://doi.org/10.1103/RevModPhys.72.1} {Rev.
  Mod. Phys. {\bf 72}, 1 (2000).}

\bibitem{jackson2001}
J. D. Jackson and L. B. Okun, Historical roots of gauge invariance,
  \href{https://doi.org/10.1103/RevModPhys.73.663} {Rev. Mod. Phys. {\bf 73},
  663 (2001).}

\bibitem{noether1918}
E. Noether, {Invariante Variationsprobleme,}
  \href{https://gdz.sub.uni-goettingen.de/download/pdf/PPN252457811_1918/LOG_0022.pdf}
  {Nachr. d. K\"onig. Gesellsch. d. Wiss. zu G\"ottingen, Math.-Phys. Klasse,
  {\bf 235}, 183 (1918).} English translation by M. A. Tavel: Invariant
  variation problems. \href{https://doi.org/10.1080/00411457108231446} {Transp.
  Theo. Stat. Phys. {\bf 1}, 186 (1971)}; for a version in modern typesetting
  see: Frank Y. Wang,
  \href{http://arxiv.org/abs/physics/0503066v3}{arXiv:physics/0503066v3}
  (2018).

\bibitem{baez2013markov}
J. C. Baez and B. Fong, {A Noether theorem for Markov processes,}
  \href{http://dx.doi.org/10.1063/1.4773921} {J. Math. Phys. \textbf{54},
  013301 (2013).}

\bibitem{marvian2014quantum}
I. Marvian and R. W. Spekkens, {Extending Noether's theorem by quantifying the
  asymmetry of quantum states,} \href{https://doi.org/10.1038/ncomms4821} {Nat.
  Commun. \textbf{5}, 3821 (2014).}

\bibitem{sasa2016}
S. Sasa and Y. Yokokura, {Thermodynamic entropy as a Noether invariant,}
  \href{http://dx.doi.org/10.1103/PhysRevLett.116.140601} {Phys. Rev. Lett.
  {\bf 116}, 140601 (2016).}

\bibitem{sasa2019}
S. Sasa, S. Sugiura, and Y. Yokokura, {Thermodynamical path integral and
  emergent symmetry,} \href{https://doi.org/10.1103/PhysRevE.99.022109} {Phys.
  Rev. E \textbf{99}, 022109 (2019).}

\bibitem{bravetti2023}
A. Bravetti, M. A. Garcia-Ariza, and D. Tapias, Thermodynamic entropy as a
  Noether invariant from contact geometry,
  \href{https://doi.org/10.3390/e25071082} {Entropy {\bf 25}, 1082 (2023).}

\bibitem{budkov2022}
Y. A. Budkov and A. L. Kolesnikov, Modified Poisson-Boltzmann equations and
  macroscopic forces in inhomogeneous ionic fluids,
  \href{https://doi.org/10.1088/1742-5468/ac6a5b} {J. Stat. Mech. {\bf 2022},
  053205 (2022).}

\bibitem{brandyshev2023}
P. E. Brandyshev and Y. A. Budkov, Noether's second theorem and covariant field
  theory of mechanical stresses in inhomogeneous ionic fluids,
  \href{https://doi.org/10.1063/5.0148466} {J. Chem. Phys. {\bf 158}, 174114
  (2023)}.

\bibitem{beyen2025clausius}
A. Beyen and C. Maes, The first part of Clausius' heat theorem in terms of
  Noether's theorem, \href{https://doi.org/10.2140/memocs.2025.13.1} {Math.
  Mech. Compl. Sys. {\bf 13}, 1 (2025).}

\bibitem{mueller2024gauge}
J. M\"uller, S. Hermann, F. Samm\"uller, and M. Schmidt, Gauge invariance of
  equilibrium statistical mechanics,
  \href{https://doi.org/10.1103/PhysRevLett.133.217101} {Phys. Rev. Lett. {\bf
  133}, 217101 (2024)}; Editors' Suggestion; PRL's
  \href{https://promo.aps.org/PRL2024} {Collection of the Year} 2024; Featured
  in \href{https://doi.org/10.1103/Physics.17.163} {Physics {\bf 17}, 163
  (2024)} by B. Rotenberg.

\bibitem{mueller2024whygauge}
J. M\"uller, F. Samm\"uller, and M. Schmidt, Why gauge invariance applies to
  statistical mechanics, \href{https://doi.org/10.1088/1751-8121/adbfe6} {J.
  Phys. A: Math. Theor. {\bf 58}, 125003 (2025).}

\bibitem{rotenberg2024spotted}
B. Rotenberg, Viewpoint: Symmetry spotted in statistical mechanics,
  \href{https://doi.org/10.1103/Physics.17.163} {Physics {\bf 17}, 163 (2024).}

\bibitem{miller2025physicsToday}
J. L. Miller, Gauge invariance applies to statistical mechanics too,
  \href{https://doi.org/10.1063/pt.wcgt.sirc} {Phys. Today {\bf 78}, 11 (2025);
  Cover image of the February 2025 issue {\bf 78} (2).}

\bibitem{mueller2024dynamic}
J. M\"uller, F. Samm\"uller, and M. Schmidt, Dynamical gauge invariance of
  statistical mechanics, \href{https://doi.org/10.48550/arXiv.2504.17599}
  {arXiv:2504.17599.}

\bibitem{hermann2021noether}
S. Hermann and M. Schmidt, {Noether's theorem in statistical mechanics},
  \href{https://doi.org/10.1038/s42005-021-00669-2} {Commun. Phys. {\bf 4}, 176
  (2021).}

\bibitem{robitschko2024any}
S. Robitschko, F. Samm\"uller, M. Schmidt, and S. Hermann, Hyperforce balance
  from thermal Noether invariance of any observable,
  \href{https://doi.org/10.1038/s42005-024-01568-y} {Commun. Phys. {\bf 7}, 103
  (2024).}

\bibitem{hermann2022quantum}
S. Hermann and M. Schmidt, {Force balance in thermal quantum many-body systems
  from Noether's theorem}, \href{https://doi.org/10.1088/1751-8121/aca12d} {J.
  Phys. A: Math. Theor. {\bf 55}, 464003 (2022). \it (Special Issue: Claritons
  and the Asymptotics of ideas: the Physics of Michael Berry)}.

\bibitem{sammueller2023whatIsLiquid}
F. Samm\"uller, S. Hermann, D. de las Heras, and M. Schmidt,
  {Noether-constrained correlations in equilibrium liquids},
  \href{https://doi.org/10.1103/PhysRevLett.130.268203} {Phys. Rev. Lett. {\bf
  130}, 268203 (2023).}

\bibitem{yvon1935}
J. Yvon, La th\'eorie statistique des fluides et l'\'equation d'\'etat (in
  French), {\it Actualit\'es Scientifiques et Industrielles}, (Hermann \& Cie.,
  Paris, 1935).

\bibitem{born1946}
M. Born and H. S. Green, {A general kinetic theory of liquids I.~The molecular
  distribution functions}, \href{https://doi.org/10.1098/rspa.1946.0093} {Proc.
  R. Soc. London, Ser. A {\bf 188}, 10 (1946).}

\bibitem{hansen2013}
J.~P. Hansen and I.~R. McDonald, {\it Theory of Simple Liquids}, 4th ed.\
  (Academic Press, London, 2013).

\bibitem{mueller2025longQuantum}
J. M\"uller and M. Schmidt, Quantum statistical mechanics: Gauge invariance,
  operator shifting, hyperdensity functionals, and nonequilibrium sum rules,
  \href{https://doi.org/10.48550/arXiv.2605.26650} {arXiv:2605.26650.}

\bibitem{bloch2008}
I. Bloch, J, Dalibard, and W. Zwerger, Many-body physics with ultracold gases,
  \href{https://doi.org/10.1103/RevModPhys.80.885} {Rev. Mod. Phys. {\bf 80},
  885 (2008).}

\bibitem{choi2016}
J.-Y. Choi, S. Hild, J. Zeiher, P. Schau\ss, A. Rubio-Abadal, T. Yefsah, V.
  Khemani, D. A. Huse, I. Bloch, and C. Gross, Exploring the many-body
  localization transition in two dimensions,
  \href{https://doi.org/10.1126/science.aaf8834} {Science {\bf 352}, 1547
  (2016).}

\bibitem{bordia2017}
P. Bordia, H L\"uschen, S. Scherg, S. Gopalakrishnan, M. Knap, U. Schneider,
  and I. Bloch, Probing slow relaxation and many-body localization in
  two-dimensional quasiperiodic systems,
  \href{https://doi.org/10.1103/PhysRevX.7.041047 } {Phys. Rev. X {\bf 7},
  041047 (2017).}

\bibitem{yan2017prl}
M. Yan, H. Y. Hui, M. Rigol, and V. W. Scarola, Equilibration dynamics of
  strongly interacting bosons in 2D lattices with disorder,
  \href{https://doi.org/10.1103/PhysRevLett.119.073002 } {Phys. Rev. Lett. {\bf
  119}, 073002 (2017).}

\bibitem{yan2017pra}
Dynamics of disordered states in the Bose-Hubbard model with confinement, M.
  Yan, H.-Y. Hui, and V. W. Scarola,
  \href{https://doi.org/10.1103/PhysRevLett.119.073002 } {Phys. Rev. A. {\bf
  95}, 053624 (2017).}

\bibitem{lydzba2023}
P. \L ydzba, M. Mierzejewski, M. Rigol, and L. Vidmar, Generalized
  thermalization in quantum-chaotic quadratic Hamiltonians,
  \href{https://doi.org/10.1103/PhysRevLett.131.060401} {Phys. Rev. Lett. {\bf
  131}, 060401 (2023).}

\bibitem{patil2026}
R. Patil and M. Rigol, Eigenstate thermalization for local versus
  translationally invariant observables,
  \href{https://doi.org/10.48550/arXiv.2602.09087} {arxiv:2602.09087.}

\bibitem{patil2026review}
R. Patil and M. Rigol, Eigenstate thermalization, in:~{\it Comprehensive
  Quantum Mechanics}, to be published by Elsevier (Main Editor: R. B. Mann;
  Volume Editors: S. Gnutzmann and K. Zyczkowski);
  \href{https://doi.org/10.48550/arXiv.2604.11872} {arXiv:2604.11872.}

\bibitem{palamara2024}
Thermal-density-functional-theory approach to quantum thermodynamics A.
  Palamara, F. Plastina, A. Sindona, I. D'Amico,
  \href{https://doi.org/10.1103/PhysRevA.110.062203} { Phys. Rev. A {\bf 110},
  062203 (2024).}

\bibitem{palamara2025}
A. Palamara, F. Plastina, A. Sindona, and I. D'Amico, Full quantum work
  statistics for non-homogeneous many-body systems,
  \href{https://doi.org/10.1088/2058-9565/ae6a1b} {Quantum Sci. Technol. {\bf
  11}, 025055 (2026).}

\bibitem{ullrich2025}
C. A. Ullrich, A snapshot of time-dependent density-functional theory,
  \href{https://doi.org/10.1063/5.0297117} {APL Comp. Phys. {\bf 1}, 020901
  (2025).}

\bibitem{schmidt2022rmp}
M. Schmidt, {Power functional theory for many-body dynamics},
  \href{https://doi.org/10.1103/RevModPhys.94.015007} {Rev. Mod. Phys. {\bf
  94}, 015007 (2022).}

\bibitem{dirac1958book}
P. A. M. Dirac, {\it The Principles of Quantum Mechanics}, 4th ed.\ (Clarendon
  Press, Oxford, 1958).

\bibitem{sakurai1973book}
J. J. Sakurai, {\it Advanced quantum mechanics} (Addison-Wesley, Reading,
  Mass., 1973).

\bibitem{fick1990book}
E. Fick and G. Sauermann, {\it The Quantum Statistics of Dynamic Processes},
  \href{https://link.springer.com/book/9783642837173} {(Springer, Berlin,
  1990)}.

\bibitem{anderson1994}
A. Anderson, Canonical transformations in quantum mechanics,
  \href{https://doi.org/10.1006/aphy.1994.1055} {Ann. Phys. {\bf 232}, 292
  (1994).}

\bibitem{goldstein2002}
H. Goldstein, C. Poole, and J. Safko, {\it Classical Mechanics}
  (Addison-Wesley, New York, 2002).

\bibitem{kubo1957}
R. Kubo, Statistical-mechanical theory of irreversible processes. I. General
  theory and simple applications to magnetic and conduction problems,
  \href{https://doi.org/10.1143/JPSJ.12.570} {J. Phys. Soc. Jpn. {\bf 12}, 570
  (1957).}

\bibitem{petz1993}
D. Petz and G. Toth, The Bogoliubov inner product in quantum statistics,
  \href{https://doi.org/10.1007/BF00739578} {Lett. Math. Phys. {\bf 27}, 205
  (1993).}

\bibitem{sauermann1996}
G. Sauermann, H. Turschner, and W. Just, Selfconsistent approximations in
  Mori's theory,
  \href{https://www.sciencedirect.com/science/article/pii/0378437195003851}
  {Physica A {\bf 225}, 19 (1996).}

\bibitem{sammueller2024hyperDFT}
F. Samm\"uller, S. Robitschko, S. Hermann, and M. Schmidt, Hyperdensity
  functional theory of soft matter,
  \href{https://doi.org/10.1103/PhysRevLett.133.098201} {Phys. Rev. Lett. {\bf
  133}, 098201 (2024); PRL Editors' Suggestion.}

\bibitem{sammueller2024whyhyperDFT}
F. Samm\"uller and M. Schmidt, Why hyperdensity functionals describe any
  equilibrium observable, \href{https://doi.org/10.1088/1361-648X/ad98da} {J.
  Phys.: Condens. Matter {\bf 37}, 083001 (2025); (Topical Review).}

\bibitem{mermin1965}
N. D. Mermin, Thermal properties of the inhomogeneous electron gas,
  \href{https://doi.org/10.1103/PhysRev.137.A1441} {Phys. Rev. {\bf 137}, A1441
  (1965).}

\bibitem{tokatly2005one}
I. V. Tokatly, Quantum many-body dynamics in a Lagrangian frame: I. Equations
  of motion and conservation laws,
  \href{https://doi.org/10.1103/PhysRevB.71.165104} {Phys. Rev. B {\bf 71},
  165104 (2005).}

\bibitem{tokatly2005two}
I. V. Tokatly, Quantum many-body dynamics in a Lagrangian frame: II. Geometric
  formulation of time-dependent density functional theory,
  \href{https://doi.org/10.1103/PhysRevB.71.165105} {Phys. Rev. B {\bf 71},
  165105 (2005).}

\bibitem{tokatly2007}
I. V. Tokatly, Time-dependent deformation functional theory,
  \href{https://doi.org/10.1103/PhysRevB.75.125105} {Phys. Rev. B {\bf 75},
  125105 (2007).}

\bibitem{ullrich2006}
C. A. Ullrich and I. V. Tokatly, Nonadiabatic electron dynamics in
  time-dependent density-functional theory,
  \href{https://doi.org/10.1103/PhysRevB.73.235102} {Phys. Rev. B {\bf 73},
  235102 (2006).}

\bibitem{tarantino2021}
W. Tarantino and C. A. Ullrich, A reformulation of time-dependent Kohn-Sham
  theory in terms of the second time derivative of the density,
  \href{https://doi.org/10.1063/5.0039962} {J. Chem. Phys. {\bf 154}, 204112
  (2021).}

\bibitem{tchenkoue2019}
M.-L. M. Tchenkoue, M. Penz, I. Theophilou, M. Ruggenthaler, and A. Rubio,
  Force balance approach for advanced approximations in density functional
  theories, \href{https://doi.org/10.1063/1.5123608} {J. Chem. Phys. {\bf 151},
  154107 (2019).}

\bibitem{daas2025}
K. J. Daas, S. Crisostomo, K. Burke, Ensemble time-dependent density functional
  theory, \href{https://doi.org/10.48550/arXiv.2507.19464} {arXiv:2507.19464}.

\bibitem{green1954}
M. S. Green, Markoff random processes and the statistical mechanics of
  time-dependent phenomena. II. Irreversible processes in fluids,
  \href{https://doi.org/10.1063%2F1.1740082} {J. Chem. Phys. {\bf 22}, 398
  (1954).}

\bibitem{deroeck2004}
W. De Roeck and C. Maes, Quantum version of free-energy--irreversible-work
  relations, \href{https://doi.org/10.1103/PhysRevE.69.026115} {Phys. Rev. E
  {\bf 69}, 026115 (2004).}

\bibitem{celeri2024}
L. C. C\'eleri and \L. Rudnicki, Gauge-invariant quantum thermodynamics:
  consequences for the first law, \href{https://doi.org/10.3390/e26020111}
  {Entropy {\bf 26}, 111 (2024).}

\bibitem{ferrari2025}
G. F. Ferrari, \L. Rudnicki, and L. C. C\'eleri, Quantum thermodynamics as a
  gauge theory, \href{https://doi.org/10.1103/PhysRevA.111.052209} {Phys. Rev.
  A {\bf 111}, 052209 (2025).}

\bibitem{nguyen2026}
N. Nguyen-Tran-Thanh, T. Nguyen-Xuan, and H. Pham-Van, Gauge theory of
  orientation in anisotropic fluids, \href{https://doi.org/10.1103/4k95-qv87}
  {Phys. Rev. E {\bf 113}, 025401 (2026).}

\bibitem{phamvan2026symmetry}
H. Pham-Van, Unified gauge-geometry symmetry for equilibrium statistical
  mechanics, \href{https://doi.org/10.1103/hvkc-djyj} {Phys. Rev. E {\bf 113},
  054134 (2026).}

\bibitem{maruyama2026}
T. Maruyama, T. Seto, V. Zaverkin, H. Christiansen, A Leibniz rule of
  distributional pairing and hyperforce sum rule,
  \href{https://doi.org/10.48550/arXiv.2603.01519} {arXiv:2603.01519.}

\bibitem{phamvan2026quantumGeometry}
H. Pham-Van, Quantum statistical-gauge geometry,
  \href{https://doi.org/10.1103/jb7b-rsfb} {Phys. Rev. E {\bf 113}, 064134
  (2026).}

\bibitem{phamvan2026exchangePrivateLindblad}
H. Pham-Van (private communication).

\bibitem{mueller2025quantumData}
Code is available at:
  \href{https://www.mschmidt.uni-bayreuth.de/pubs/quantum_gauge.nb}
  {https://www.mschmidt.uni-bayreuth.de/pubs/quantum\_gauge.nb}.

\end{thebibliography}

\end{document}